\documentstyle[12pt,preprint,aps,epsf,floats]{revtex}

\tighten

\begin{document}

\preprint{
\noindent
\hfill
\begin{minipage}[t]{3in}
\begin{flushright}
CERN--TH/99--118     \\
PM/98--38            \\
RU--98--45           \\
hep-ph/9905443       \\
\vspace*{1.5cm}
\end{flushright}
\end{minipage}
}

\draft

\title{Charged-Higgs- and $R$-Parity-Violating 
Slepton-Strahlung \\ at Hadron Colliders}
\author{Francesca Borzumati$^{a}$, Jean-Lo\"\i c Kneur$^{b}$,
        and Nir Polonsky$^{c}$}
\address{${}^{a}$
 Theory Division, CERN, CH-1211 Geneva 23, Switzerland}
\address{${}^{b}$ 
 Laboratoire de Physique Math\'ematique et Th\'eorique, 
 Universit\'e de Montpellier II, \\  F--34095 Montpellier Cedex 5, France}
\address{${}^{c}$
 Department of Physics and Astronomy, 
 Rutgers University, \\ Piscataway, NJ 08854-8019, USA} 

\maketitle

\begin{abstract}

It is shown that the radiation of a charged Higgs boson off a
third-generation quark (charged-Higgs-strahlung) provides an
important channel for the discovery of the charged Higgs at hadron
colliders.  Equivalently, in supersymmetric models with explicit
lepton-number ($R$-parity) violation, sleptons may also be produced
in association with quarks (slepton-strahlung).  Higgs-- and
slepton-strahlung production cross sections are given for both the
Tevatron and the LHC.  The LHC cross sections imply that heavy
${\cal{O}}$(TeV) charged Higgs bosons can be produced via
charged-Higgs-strahlung and that strahlung production of charged
sleptons is possible even for small $R$-parity violating
couplings. The possible discovery of sleptons through this channel
offers a surprising handle on models of neutrino masses.

\end{abstract}


\newpage 
\setlength{\parskip}{1.2ex}

\section{Introduction}
\label{sec:s1}

While the LEP experiments reach their final phases, the upgraded
Tevatron ($\sqrt{s} = 2\,$TeV) at FNAL and the CERN Large Hadron
Collider (LHC) ($\sqrt{s} = 14\,$TeV) will provide the next front in
searches for particles associated with physics beyond the Standard
Model (SM).  In particular, since LEP cannot probe masses
significantly above the $100\,$GeV mark, the probe of heavier
particles will remain the task of these two hadronic machines.
Sufficient production cross sections at these colliders, however, may
be difficult to achieve if the new particles carry only weak charges.
Effective couplings of these particles to (initial) gluons may be
induced at the quantum level, but they suffer from loop suppression
and, hence, are generically small.  Sufficiently large and measurable
production cross sections require either some enhancement of the
radiative gluonic couplings or an alternative production mechanism.

In the case of the Higgs bosons present in supersymmetric models, as
well as non-supersymmetric two-Higgs-doublet models, there exist two
potentially large parameters that can partially compensate for the
otherwise small couplings: the intrinsically large $t$-quark Yukawa
coupling and the ratio of vacuum expectation values (vev's) of
the two neutral Higgs bosons $\tan \beta = v_2/v_1$, which is
constrained from above to be $\lesssim 60$ by the perturbativity of
Yukawa couplings.  They may $(i)$ enhance the radiatively induced
gluonic coupling of the neutral Higgs, in proportion to the $t$-quark
Yukawa coupling, leading to production via 
$gg \to H^{0}$~\cite{FUSION}; $(ii)$ sufficiently increase 
the rate for the decay of the $t$-quark into a (light) charged Higgs
boson, ${t}\to H^{+}{b}$~\cite{BD,TDECAYlast,TDECAY,TDECAYold}; 
or $(iii)$ enhance the Higgs-strahlung associate production through
the $2\to 3$ partonic processes
$gg,\, qq\to qqH^{0},\,qq^{\prime}H^{\pm}
$~\cite{TTH,TBH,JGUNION,DW,BBH}, 
through the $2 \to 2$ ones 
$qg \to qH^{0},\,q^{\prime}H^{\pm}$~\cite{DW,BG,BGII}, 
and through the $2 \to 1$ process $\bar{b} b \to H^0$~\cite{DW,BBH}.
(For an overview, see, for example, Ref.~\cite{GDRMSSM}.)

{}From charge conservation, a radiatively induced gluonic coupling
cannot lead, at the level of an elementary process, to the production
of only one charged Higgs.  Hence, one needs to consider either the
single production from $t$-quark decays or the production in
association with quarks -- which is referred to as Higgs-strahlung.
Production in association with gauge bosons $W H^0$/$Z H^0$,
$W^{\pm}H^{\mp}$, and in supersymmetric models, in association with
squarks, is also possible. The former mechanism, however, leads to
subleading cross sections, which are difficult to
observe~\cite{WH,WHPLUS}.  The latter~\cite{STOPS,STOPS2,SQUARKS} is more
model-dependent and will not be considered here.  The production
mechanisms with one single Higgs (and quarks) in the final state, or
``single production'', are kinematically in advantage with respect to 
pair-production mechanisms. The latter includes $(a)$ the
Drell--Yan process $q\bar{q} \to H^{+}H^{-}$, which is suppressed by
weak couplings, and, at the LHC, by the low quark luminosities (relative to
the gluon one); $(b)$ effective gluonic couplings, which are now
allowed by charge invariance, $gg \to H^{+}H^{-}$~\cite{PAIR}.  Pair
production, therefore, does not allow the discovery of the charged Higgs 
at the Tevatron. At the LHC, it provides a limited discovery reach of 
the charged Higgs in a non-supersymmetric two-Higgs-doublet
model~\cite{PAIR}, but it may become more competitive in supersymmetry
where additional contributions to the effective gluonic couplings
arise~\cite{BP,MD}.

Our focus here is on the single production of a charged Higgs boson.
Single production from $t$-quark decays plays the most important role
when it is kinematically allowed, that is for
$m_{H^{\pm}}<m_{t}-m_{b}$, and was studied extensively by various
authors~\cite{TDECAY}.  The charged-Higgs-strahlung in the $2\to 3$
channels $gg,\, qq \to t H^- \bar{b}$ encompasses the resonant
production of a pair $t\bar{t}$ followed by the decay 
$\bar{t} \to H^-\bar{b}$ in the same kinematical region 
$m_{H^{\pm}}< m_{t}-m_{b}$; in addition, it provides the intrinsically
off-shell associate production of $H^\pm$ beyond this kinematical
limit, throughout all possible ranges of $m_{H^\pm}$.  Strahlung from
the $2\to 2$ channel $g b \to t H^-$ is also possible, since the
$b$-quark is obtained from the proton via a gluon. Therefore, both
types of partonic processes, the $2 \to 3$ and the $ 2 \to 2$, give
rise to inclusive processes that are formally of the same order in an
$\alpha_s$ expansion.  Note that away from the resonance region, $H^-$
decays dominantly into $\bar{t} b$, and the $t$-quark into $W^+b$.
Thus, the $2 \to 3$ processes $gg,\, qq \to t H^- \bar{b}$ give rise
to four $b$-quarks in the final state, whereas only three $b$'s are
produced in the $2 \to 2 $ process $g b \to t H^-$.  Both elementary
processes contribute to the inclusive production of a single charged
Higgs at hadron colliders, when at most three $b$-quarks are tagged
and used to identify the final particle configuration. If four $b$'s
can be detected, it is possible to measure each cross section
separately. Otherwise, the two elementary processes have to be
properly combined into an inclusive cross section, avoiding double
counting of the contributions coming from $g b \to t H^-$ and from 
$gg \to t H^{-}\bar{b}$, when one of the two gluons produces a
$b\bar{b}$ pair collinear to the initial proton (or antiproton).
Since identification and detection issues will not be discussed here,
both the individual-channel and inclusive cross sections will be given
below.

The charged-Higgs production cross sections are calculated and
illustrated for the upgraded Tevatron and the LHC in
Section~\ref{sec:s2}, where all relevant issues are discussed in
detail.  Attention is given to the prospects of discovery at the
Tevatron, which may constrain $m_{H^\pm}$ beyond the kinematical limit
$m_{H^{\pm}} < m_{t}-m_{b}$. On the LHC front, 
it is found that a charged Higgs as
heavy as ${\cal{O}}$(TeV) may be produced at the LHC via the strahlung processes.
The results shown are valid for the charged Higgs of
supersymmetric and non-supersymmetric two-Higgs-doublet models. The
charged-Higgs decay modes, however, may differ in the two classes of
models if $m_{H^\pm}$ is sufficiently large~\cite{TDECAYlast,BD}.

{}For reference, the indirect lower limit on $m_{H^\pm}$ coming from
the measurement of the inclusive decay $b\to s \gamma$ amounts (at
present) only to $\sim 165\,$GeV~\cite{INDIRECTLIMITS} in
non-supersymmetric two-Higgs-doublet models.  No substantial limit
exists for charged Higgs in supersymmetric models, when supersymmetric
partners can be exchanged in the loop mediating the $b \to s \gamma$
decay.  Direct lower bounds on $m_{H^{\pm}}$ are given by collider
searches at LEP II and at the Tevatron. The LEP II bound, 
$m_{H^{\pm}} \gtrsim 54\,$GeV~\cite{PDG} at $\sqrt{s} = 130\,$GeV
($m_{H^{\pm}}\gtrsim 68\,$GeV at higher energy runs~\cite{GDRMSSM}),
applies to the case of a charged Higgs boson present in
two-Higgs-doublet models.  The Tevatron searches give combined (and
currently modest) bounds in the $m_{H^{\pm}}$--$\tan \beta$ plane for
supersymmetric and non-supersymmetric charged Higgs bosons.  Both
searches have been critically discussed in
Ref.~\cite{BD}. Conservatively, all results presented in this paper
are shown for $m_{H^{\pm}} \gtrsim 45\,$GeV, the model-independent
limit extracted from the measurement of the $Z$ width.

In supersymmetric models in which $R$-parity and lepton number are not
conserved, the hypercharge $Y = -1$ Higgs and slepton fields are not
distinguished by any quantum numbers and could interact in a similar
way.  Thus, in these models, sleptons can be produced via
slepton-strahlung just as the (charged) Higgs. The relevant Yukawa
couplings, i.e.  the slepton--fermion--fermion couplings, are subject
to various low-energy constraints but are otherwise arbitrary, since
they do not relate to fermion masses (once the two Higgs doublets are
identified as those whose neutral components are aligned along the two
large vev's).  They are, however, related to radiative neutrino
masses, as explained below, and therefore measurements of
lepton-number-violating operators from slepton production provide a
unique and important window on this class of models for neutrino
masses.  The associate production cross section for the charged
sleptons is presented for both the Tevatron and the LHC experiments in
Section~\ref{sec:s3}. It is shown that charged sleptons can be
produced in abundance for $R$-parity-violating couplings as small as
$0.01$. Relations and possible lessons to models of neutrino masses
are also demonstrated in Section~\ref{sec:s3}.  We also comment on the
case of the neutral sleptons, the sneutrinos $\tilde{\nu}$, which is
complicated by the presence of the gluon fusion channel 
$gg \to \tilde{\nu}$.

Results and discussions of the potential impact of these strahlung
channels on future searches are summarized in Section~\ref{sec:s4},
where we also comment on the possibility that both charged-Higgs and
slepton-strahlung channels are present.

All calculations are done at the leading order in QCD.  Higher-order
corrections may be important, as was shown in the case of
associate production of the neutral Higgs~\cite{HIGGS}.  Their
inclusion is called upon, but this is left for future study.

\section{Charged-Higgs-Strahlung}
\label{sec:s2}

The charged Higgs boson interacts with quarks according to the
Lagrangian
\begin{equation} 
{\cal L} \ = \  \frac{\,g}{\sqrt{2}} \, \left\{ 
 \left(\frac{{m_d}_i}{M_W} \tan \! \beta \right) V_{ji} \,  
       {\overline{u}_L}_j \, {d_R}_i  +
 \left(\frac{{m_u}_i}{M_W} \cot \! \beta \right) V^\ast_{ji} \, \,
       {\overline{u}_R}_i \, {d_L}_j 
                                   \right\} H^+ +{\rm h.c.}\,, 
\label{higgslag}
\end{equation}
where the standard notations for the SU(2) coupling $g$, the up- and
down-quarks $u_{i}$ and $d_{i}$ of a generation $i$, and for the
Cabibbo--Kobayashi--Maskawa (CKM) matrix $V$ are used.  Hereafter,
all intergenerational mixing terms are neglected, as well as all
Yukawa couplings other than those for the $t$- and $b$-quarks, which
have respectively strength
$h_t \simeq g m_t /(\sqrt{2} M_W \sin\beta) \simeq 1/\sin\beta$ 
and
$h_b \simeq g m_b /(\sqrt{2} M_W \cos\beta) \simeq 0.017\tan\beta$.
(The current mass $m_{b} = 3$ GeV is used hereafter.
Note that using the pole mass instead 
will increase $h_{b}$, an affect that is important
in the large $\tan\beta$ regime.)
Since all calculations are done to leading order, model-dependent
radiative corrections to these relations~\cite{TDECAY,MB} are also
omitted. They could, however, be large and play an important role 
(at the order in perturbation theory at which they must be included) 
by smearing the $\tan\beta$ dependence of the $H^\pm$-production cross
sections.

\begin{figure}[th]
\begin{center}
\epsfxsize= 8.9cm
\leavevmode
\epsfbox[150 595 480 705]{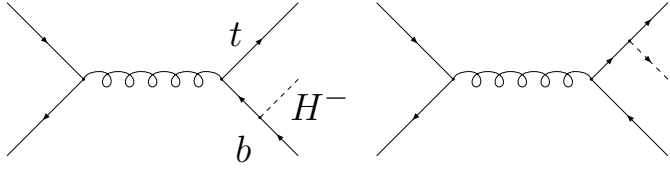}
\end{center}
\caption[f1]{Diagrams contributing to 
 $p \bar{p}\,(pp) \to t H^-(\bar{b})X$ through an elementary
 quark-initiated $2\to 3$ process.}
\label{qdiags}
\end{figure}

\begin{figure}[t]
\begin{center}
\epsfxsize= 11.5cm
\leavevmode
\epsfbox[100 480 530 705]{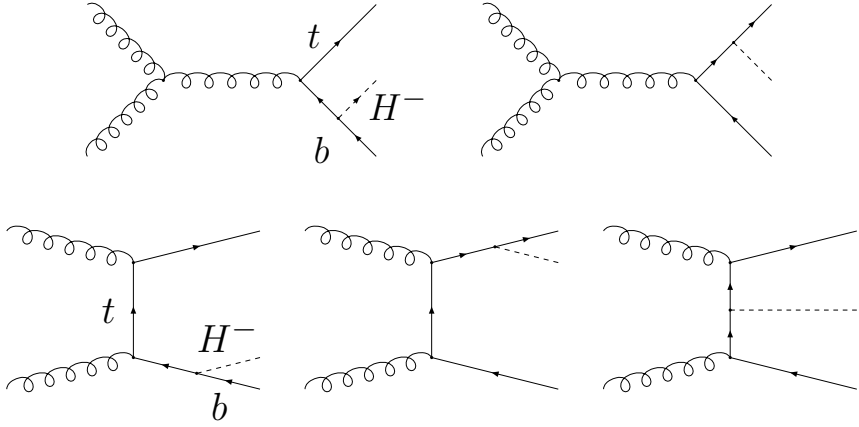}
\end{center}
\caption[f1]{Diagrams contributing to 
 $p \bar{p}\,(pp) \to t H^{-}(\bar{b})X$ through an elementary
 gluon-initiated $ 2\to 3$ process.}
\label{gdiags}
\end{figure}

\subsection{The $2 \to 3 $ process: cross-section calculation}
\label{twotothree}

Quarks produced in quark or gluon collisions can radiate a Higgs boson
leading to the associate production $gg,\,q\bar q\to t H^{-}\bar{b}$.
Diagrams describing the quark-initiated parton processes are shown in
Fig.~\ref{qdiags}, those describing the gluon-initiated processes in
Fig.~\ref{gdiags}.  As explained in the introduction, if the
center-of-mass energy of the relevant collider is sufficiently large
to allow the tagging of two $b$-quarks in addition to the
$H^\pm$-decay products, the measurement of the $2\to 3$ production
cross section may be possible.

The production cross section for a $p \bar{p}$ ($p p$) collider is
obtained as usual by convoluting the hard-scattering cross sections of
the quark- and gluon-initiated processes with the 
quark- and gluon-distribution functions inside $p$ and/or $ \bar {p}$:
\begin{eqnarray}
 \sigma                  &   &  \ = \ 
 \frac{1}{2 s} \int^1_{\tau_{\rm min}} \frac{d \tau }{\tau}  
\int^1_{\tau} \frac{d x}{x} \,
 \int d {\rm PS} (q_1+\!q_2;p_1,\!p_2,\!p_3) \times 
    \nonumber \\         &   &          
\hspace*{1truecm}
\left\{ \sum_{q}
\Big[ q(x,\mu_f) \bar{q} (\tau/x,\mu_f) + q \leftrightarrow \bar{q}
\Big]  \vert {\cal M} \vert^2_{q \bar{q}}          \,  + \,
      g(x,\mu_f) g(\tau/x,\mu_f)  
       \vert {\cal M}\vert^2_{gg}  
\right\} \,,                                         
\label{Xsec}
\end{eqnarray}
where $q_{i}$ ($p_{i}$) are the four-momenta of the initial partons
(final-state particles), the functions $q(x,\mu_f)$,
$\bar{q}(x,\mu_f)$ and $g(x,\mu_f)$ designate respectively the quark-,
antiquark- and gluon-density functions with momentum fraction $x$ at
the factorization scale $\mu_f$, and the index $q$ in the sum, runs
over the five flavors $u, d, c, s, b$. Finally,
$s$ is the hadron center-of-mass energy squared, whereas the parton
center-of-mass energy squared is indicated, as usual, by 
$\hat{s}= x_1 x_2\, s \equiv \tau s$, with 
$\tau_{\rm min}=(m_{H^\pm} +m_t +m_b)^2/s$. The integration over
$d$PS$(q_1+q_2;p_1, p_2, p_3)$, an element of phase space of the 
3-body final state, can be reduced to four nested integrals with bounds
explicitly given in appendix~\ref{dphi3}.

When rewriting the third-generation vertex $t$--$b$--$H^\pm$ 
in~(\ref{higgslag}) as
$i (g/\sqrt{2})\,V_{tb}\left(v+a\,\gamma_5\right)+{\rm h.c.}$,  
with vector and axial couplings $v $ and $a$ given by:  
\begin{equation}
 v  =   \frac{1}{2}
  \left(\frac{m_b}{M_W} \tan \! \beta + 
        \frac{m_t}{M_W} \cot \! \beta  \right)\,;   
\hspace*{1truecm}
 a  =   \frac{1}{2}
  \left(\frac{m_b}{M_W} \tan \! \beta - 
        \frac{m_t}{M_W} \cot \! \beta  \right) \,,
\label{vavalues}
\end{equation}
the square amplitudes 
$\vert {\cal M} \vert^2_{q \bar{q}}$ and 
$\vert {\cal M} \vert^2_{gg}$ 
can be decomposed as:
\begin{eqnarray}
\vert {\cal M} \vert^2_{q \bar{q}}  & = & 
 \left(4 \frac{G_F}{\sqrt{2}} M_W^2\right) \left( 4\pi \alpha_S\right)^2 
 \vert V_{tb}\vert^2 
\left(v^2 \, V^{q \bar{q}}  + a^2 \, A^{q \bar{q}}\right)
  \nonumber \\
\vert {\cal M} \vert^2_{gg}        & = &
 \left(4 \frac{G_F}{\sqrt{2}} M_W^2\right) \left( 4\pi \alpha_S\right)^2 
 \vert V_{tb}\vert^2 
\left( v^2 \, V^{gg} + a^2 \, A^{gg} \right)\,.
\label{amplitdecomp}
\end{eqnarray}
In this notation color factors are included in the reduced squared
amplitudes $V^{q \bar{q}}$, $ A^{q \bar{q}}$ and $V^{gg}$, $ A^{gg}$,
while the strong coupling constant is explicitly factored out.  The
expressions for the reduced amplitudes are too lengthy to be given
here\footnote{The Fortran code for these amplitudes is available upon
 request.}.

\begin{figure}[p]
\begin{center}
\epsfxsize= 11.7cm
\leavevmode
\epsfbox[25 165 585 565]{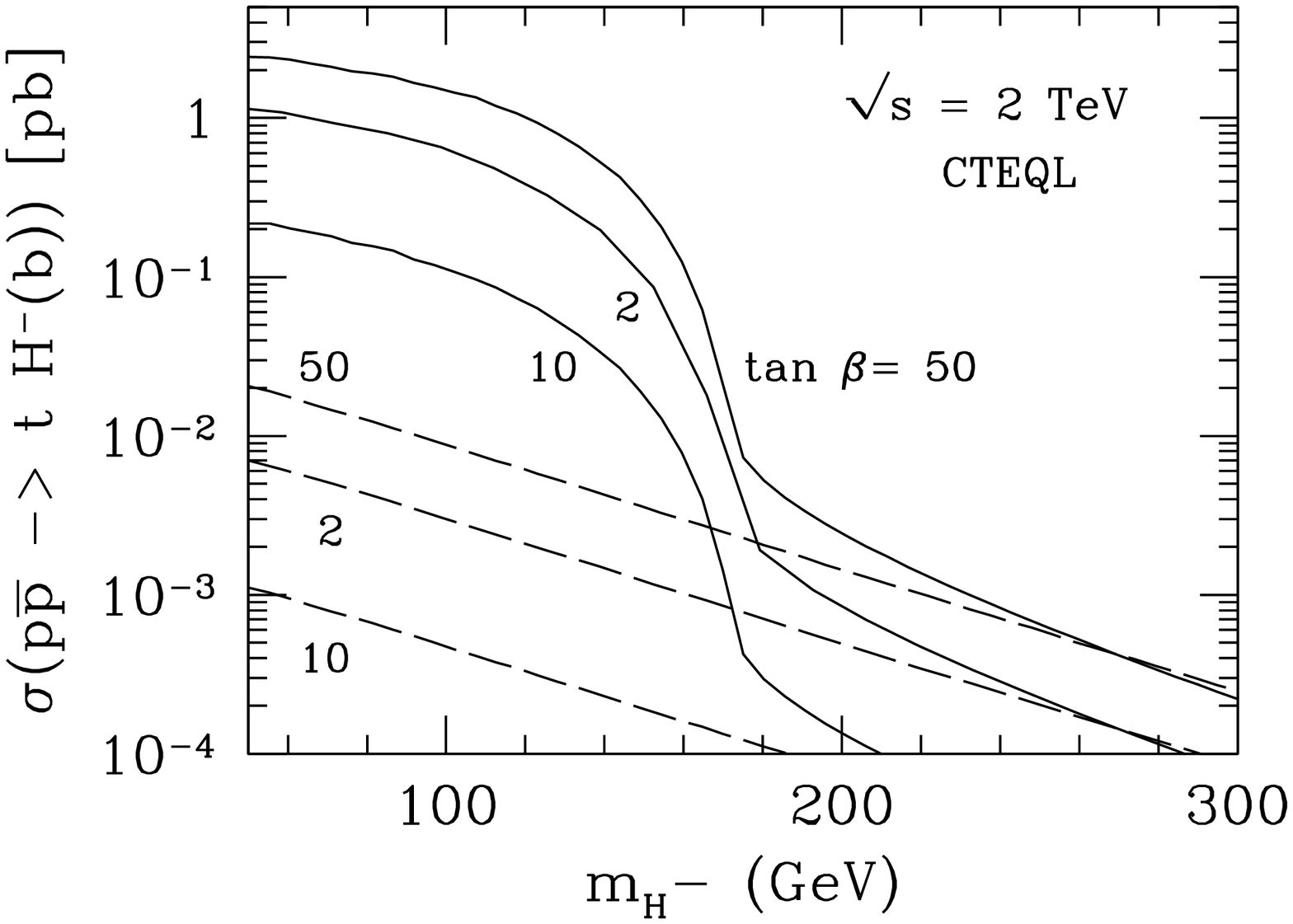}
\end{center}
\caption[f1]{The leading-order production cross section
 $\sigma(p\bar{p}\to t(\bar{b})H^- X)$ as a function of 
 the charged Higgs mass,  
 for $\sqrt{s} = 2\,$TeV, is shown for three different values of 
 $\tan\!\beta = 2,\,10,\,50$.  The solid lines indicate the cross
 sections obtained from $2\to 3$ elementary processes
 $g g \to t \bar{b} H^-$ and $q\bar{q} \to t \bar{b} H^-$; the dashed 
 lines show the cross sections obtained from the $2 \to 2$ process 
 $g b \to t H^-$. Renormalization and factorization scales are fixed
 as $\mu_R\! =\! \mu_f\! =\! m_t+m_{H^-}$.}
\label{TeVmassdep}
\vspace*{0.5truecm}
\begin{center}
\epsfxsize= 11.7cm
\leavevmode
\epsfbox[25 165 585 565]{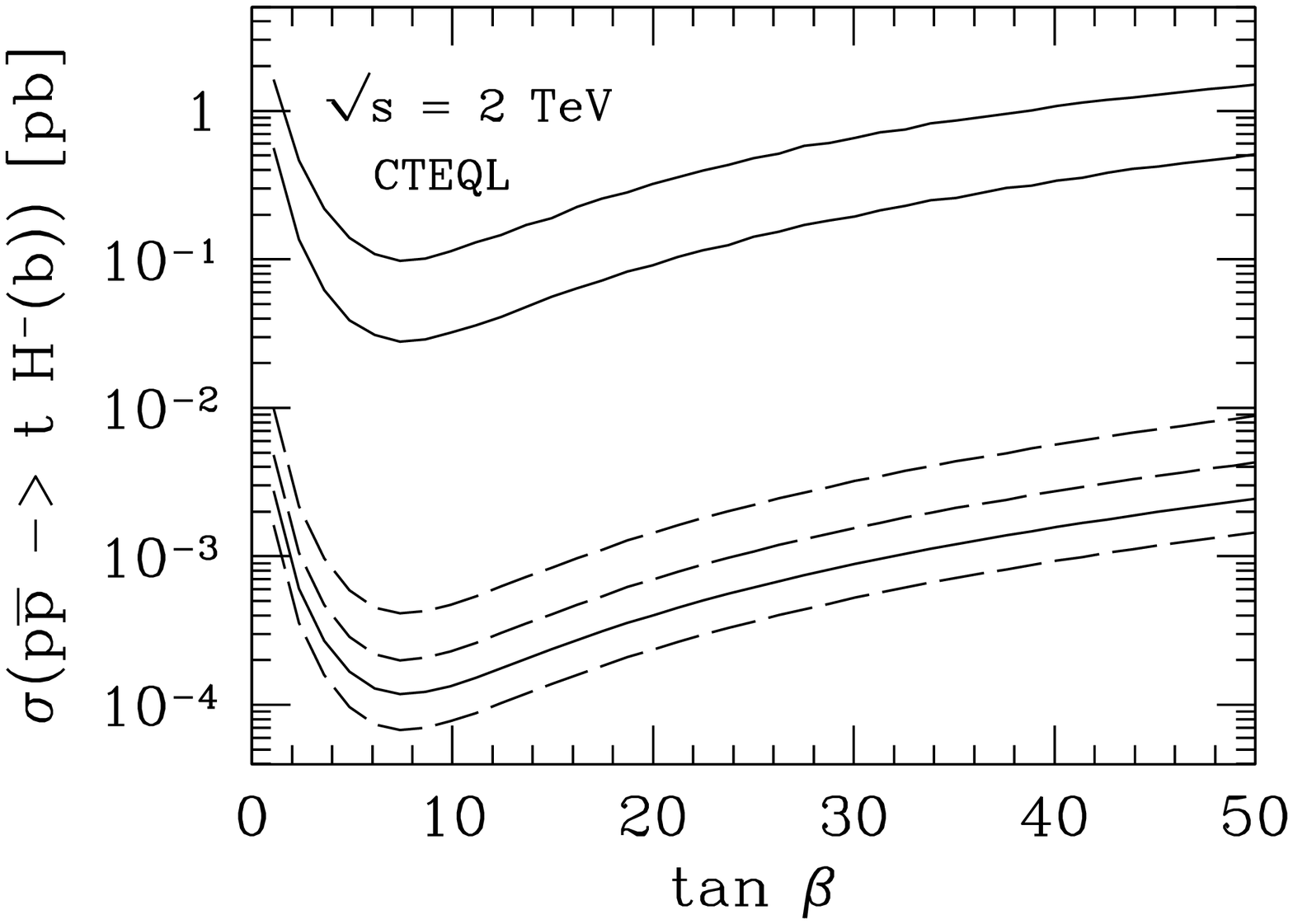}
\end{center}
\caption[f1]{The production cross section
 $\sigma(p\bar{p}\to t(\bar{b})H^- X)$ is shown as a function of
 $\tan\!\beta$ for $m_{H^\pm}=100$, $140$, and $200\,$GeV (from top 
 to bottom). As in Fig.~\ref{TeVmassdep}, solid lines correspond to
 the $2\to 3$ processes, dashed lines to the $2 \to 2$ process.}
\label{TeVtanbdep}
\end{figure}

In the kinematical region $m_t > m_{H^\pm} +m_b$ the above cross
section could be well approximated by the much simpler resonant
production cross section, as given by the on-shell $t \bar t$
production cross section times the branching fraction for the
decay $\bar{t} \to H^- \bar{b}$.

\subsection{The $2 \to 3 $ process: numerical results}
\label{twotothreeresul}

The production cross section $\sigma(p \bar{p} \to t H^- \bar{b} X)$
for the Tevatron is shown by the solid lines in Fig.~\ref{TeVmassdep}
as a function of the charged-Higgs mass for three different values of
$\tan\beta$, $\tan \beta =2$, $10$, and $50$, and in
Fig.~\ref{TeVtanbdep} as a function of $\tan\beta$ for different
values of $m_{H^{+}} = 100$, $140$, and $200\,$GeV.  Notice that these
figures show the cross section for single production of $H^-$ only and 
that identical results are obtained for the process 
$\sigma(p \bar{p} \to \bar{t} H^+ b X)$.  All calculations are done at
the leading order in QCD.  The leading-order parton distribution
functions CTEQ4L~\cite{CTEQ} are used in all calculations, and the
renormalization ($\mu_R$) and factorization ($\mu_f$) scales are
always fixed to the threshold value $m_t + m_{H^\pm}$.  The variation
of these scales results in general in changes in the cross section
presented here: a variation in the interval between 
$(m_t + m_{H^\pm})/2$ and $2(m_t + m_{H^\pm})$ can produce deviations
up to $\pm 30\%$ with respect to the values shown in the figures.
Higher-order corrections, therefore, may be important, as was shown in
the case of associate production of the neutral Higgs.  Their
inclusion is called upon, but it is left for future work.
As a cross-check of our cross-section calculation and phase-space
integration procedure, we have reproduced, by taking the appropriate
limit, the well-known $gg, q \bar q \to t \bar t h$ 
cross section~\cite{TTH} for the neutral Higgs $h$ production in 
association with $t$-quarks.

\begin{figure}[t]
\begin{center}
\epsfxsize= 11.7cm
\leavevmode
\epsfbox[25 165 585 585]{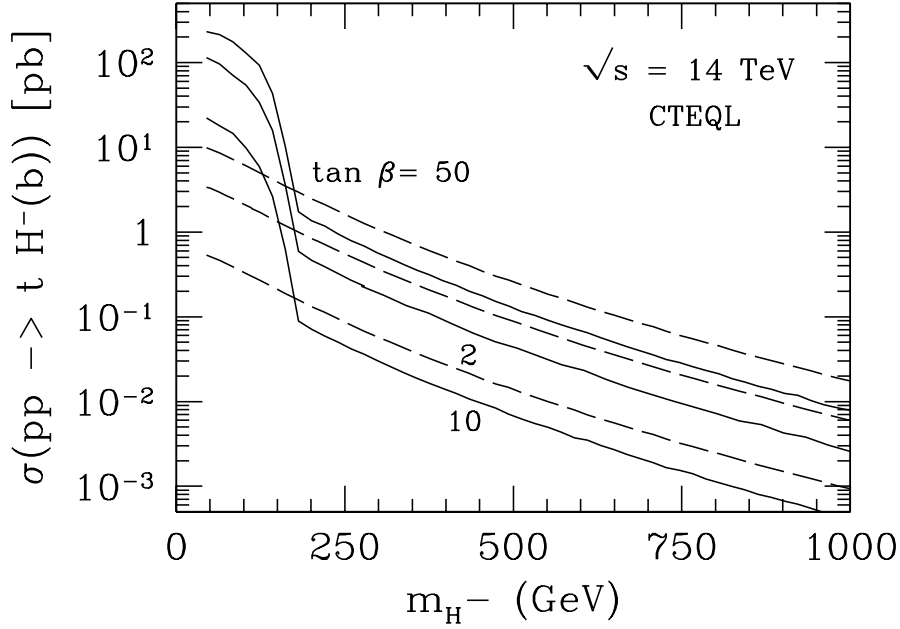}
\end{center}
\caption[f1]{The leading-order production cross section
 $\sigma(pp\to t (\bar{b}) H^- X)$, for $\sqrt{s} = 14\,$TeV
 is shown  as a
 function of the charged Higgs mass for three different values of
 $\tan \!\beta = 2,\,10,\,50$. As in
 Figs.~\ref{TeVmassdep} and~\ref{TeVtanbdep}, the solid lines indicate
 the cross sections obtained from $2\to 3$ elementary processes,
 the dashed lines, the cross sections obtained from the
 $2 \to 2$ process. Renormalization and factorization scales 
 are fixed as $\mu_R\! =\! \mu_f\! =\! m_t+m_{H^-}$.}
\label{LHCmassdep}
\end{figure}

As expected, the production cross section is enhanced in the resonance
region $m_{H^{\pm}} < m_{t}-m_{b}$, when $H^\pm$ is obtained as a
decay product of one of two $t$-quarks produced on shell (see first
diagram in Fig.~\ref{qdiags} and the first and third diagrams in
Fig.~\ref{gdiags}). The resonant $t$-quark propagator is regularized
by the width of the $t$-quark, calculated from the SM decay 
$t\to b W^+$:
\begin{eqnarray}
 \lefteqn{\Gamma(t\to b W^+)  =  } \nonumber \\
                                    & &   
 \frac{g^2}{64 \pi} 
 \vert V_{tb}^\ast \vert^2  
 \frac{1}{m_t \,M_{W^+}^2}  
 \left[M_{W^+}^2 \left(m_t^2 +m_b^2\right) + 
                 \left(m_t^2 -m_b^2\right)^2 - 2 M_{W^+}^4
 \right]
 \lambda^{1/2}\left(1,\frac{m_{W^+}^2}{m_t^2}, 
                               \frac{m_b^2}{m_t^2}\right) \,,
\end{eqnarray}
where $\lambda$ is the K\"allen function
$\lambda(x,y,z) = ((x^2 -y^2-z^2)^2 - 4 yz)$; 
and from the decay ${t} \to  H^{+}{b}$:
\begin{eqnarray}
 \lefteqn{\Gamma({t}\to H^+ {b})  =  } \nonumber \\
                                    & &   
 \frac{g^2}{32 \pi} 
 \vert V_{tb}^\ast \vert^2 \,m_t   
 \left\{ v^2 \left[ \left( 1\!+\! \frac{m_b}{m_t} \right)^2 
                            \!-\! \frac{m_{H^+}^2}{m_t^2} \right] 
     +   a^2 \left[ \left( 1\!-\! \frac{m_b}{m_t} \right)^2 
                            \!-\! \frac{m_{H^+}^2}{m_t^2} \right] 
 \right\} \lambda^{1/2}\left(1,\frac{m_{H^+}^2}{m_t^2}, 
                               \frac{m_b^2}{m_t^2}\right)\,,
\end{eqnarray}
for each value of $m_{H^\pm}$ and $\tan \!\beta$. 
In this region the cross section is not distinguishable from the
convolution of $\sigma (p \bar{p} \to t \bar{t}X)$ with the branching
fraction $Br({t} \to H^{+}{b})$.  The $\tan\beta$ dependence, shown
explicitly in Fig.~\ref{TeVtanbdep}, has the same typical pattern as
the branching ratio $Br({t} \to H^{+}{b})$, i.e. large enhancements
for very small and very large values of $\tan \! \beta$ and a minimum
for $\tan\beta \simeq \sqrt{m_t/m_b}$.  Away from the resonance
region, the cross section diminishes rather rapidly and becomes
negligible (at the Tevatron energy) for 
$m_{H^{\pm}} \gtrsim 250\,$GeV.

Results qualitatively similar to those found for the Tevatron
center-of-mass energy are obtained in the case of LHC searches.  They
are shown in Fig.~\ref{LHCmassdep}.  Assuming integrated luminosity of
100 fb$^{-1}$, a significant production cross section is obtained for
$m_{H^\pm} \lesssim 1\,$TeV, even for $\tan\beta = 10$, which is near
the minimum of the production cross section.

Our results generalize those of Ref.~\cite{TBH}, where contributions
from the gluon-initiated diagrams were calculated for the LHC but only
for $m_{t} = 150\,$GeV and for a fixed $\tan\beta =1$. Both of these
values are now experimentally ruled out. We also note a disagreement
between our calculation and that of Ref.~\cite{TBH}, which could be,
in part, due to our usage of a current set of structure functions.  A
similar calculation for the LHC case was also presented in
Ref.~\cite{JGUNION}, for different values of $m_t$.  No immediate
comparison with our results is, however, possible.  The mechanism of
associate production was also emphasized in Ref.~\cite{SOLA}, and a
preliminary study was presented for the Tevatron center-of-mass
energy~\cite{SOLAtalk}.

\subsection{The $2 \to 2 $ process}
\label{twototwo}

\begin{figure}[ht]
\begin{center}
\epsfxsize= 8.5cm
\leavevmode
\epsfbox[150 595 460 710]{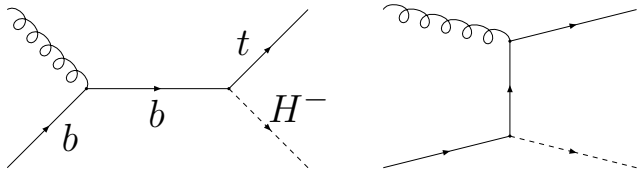}
\end{center}
\caption[f1]{Diagrams contributing to 
 $p \bar{p}\,(pp) \to t H^- X $ through an elementary
 $2 \to 2$ process.}
\label{gbdiags}
\end{figure}

As mentioned in the introduction, for an inclusive measurement of
single production of $H^-$, or if only one $b$-quark can be tagged
besides the decay products of $H^-$, the elementary process 
$g b \to t H^-$ has to be considered as well. The corresponding diagrams
are shown in Fig.~\ref{gbdiags}, and the hard-scattering cross section
reads:
$$
\sigma (g\,b \to H^-\, t) = 
\left(4\frac{G_F}{\sqrt{2}} M_W^2\right)  
\left(4\pi\alpha_s\right)  \vert V_{tb}\vert^2 
\frac{1}{192 \pi\hat{s} \left(1\!-\!x^2_b \right)^3} 
\left\{8 C_- x_b x_t 
\left[L\,(1\! -\!x^2_{ht})-2\,b\right] +  
\right. 
$$
\begin{equation}
\left.  C_+
\left[ 2L
\left(1+\!x^4_b- \!2 x^2_b x^2_{ht} -\!2 x^2_{ht} 
     (1\!-\!x^2_{ht})\right) 
- b
\left(3-\!7 x^2_{ht}+\!x^4_b (3\!+\!x^2_{ht})+\!2x^2_b 
     (1\!-\!x^2_{ht})\right)  
\right]   \right\}\,,
\label{siggb}
\end{equation}
where 
$C_\pm = v^2 \pm a^2$;  
$x_i \equiv m_i/\sqrt{\hat s}$, $x^2_{ht} = x^2_h -x^2_t$; 
$b = \left[ (1-(x_t+x_h)^2)(1-(x_t -x_h)^2) \right]^{1/2}$, and 
$L \equiv 
\ln\left[(1-x^2_{ht} +b) /(1-x^2_{ht} -b)\right] 
$.

The cross section originated from the $2 \to 2 $ process only is shown
in dashed lines in Figs.~\ref{TeVmassdep} and~\ref{TeVtanbdep} for the
upgraded Tevatron and in Fig.~\ref{LHCmassdep} for the LHC.  (We note
a disagreement with Ref.~\cite{BGII}.)  All calculations are again
done at the leading order in QCD, and renormalization and
factorization scales are fixed as $\mu_R=\mu_f=m_{H^\pm}+m_t$. The
cross sections are plagued by the same large uncertainties due to
scale variations already observed in the case of the $2 \to 3$
processes.  Notice that, away from the resonance region, 
$m_{H^\pm} <m_t -m_b$, the relative size of the two classes of cross
section depends on $\sqrt{s}$ and $m_{H^\pm}$.  Indeed, at high
energies, where the gluon-initiated $2 \to 3$ processes dominate over
the quark-initiated ones, the cross section arising from the
elementary process $g b \to t H^-$ is larger than that from $2 \to 3$
ones, which is penalized by a 3-body phase space suppression.  At the
Tevatron center-of-mass energy, the quark-initiated $2 \to 3$
processes still have the dominant role for values of $m_{H^\pm}$ that are
not too large.  For the particular choice of scales $\mu_R$ and
$\mu_f$ made here, the cross-over for the two cross sections is at
about $m_{H^\pm} \sim 265\,$GeV.

\subsection{The inclusive single-$H^{\pm}$ production cross section}

Before presenting the inclusive cross section, some elaboration on the
summation procedure used to add the $2\to 3$ and $2\to 2$ channels is
in order.  Since the initial $b$-quark is contained in the proton or
antiproton via a gluon, the $2\to 2$ process is of the same order in
$\alpha_s $ as the $2 \to 3$ ones. The collinearity of the $b$-quark
with the initial gluon induces the large factor
$\alpha_s(\mu_R)\log(\mu_f/m_b)$, where the factorization scale
$\mu_f$ is ${\cal{O}}(m_{H^\pm})$ and it was chosen to be 
$\mu_f = m_t+m_{H^\pm}$ in our numerical evaluations.
This factor is then resummed to all orders 
$(\alpha_s(\mu_R) \log(\mu_f/m_b))^n$ when making use of the
phenomenological $b$-distribution function.  The first order $n=1$, is
also contained in the set of $2\to 3$ partonic processes $gg \to t H^-
\bar{b}$ when one of the two initial gluons produces a pair $b
\bar{b}$ collinear to the initial $p$ or $\bar{p}$.  (See the last two
diagrams of Fig.~\ref{gdiags}.)  Thus, when summing the contributions
from $2 \to 2$ and $2 \to 3$ cross sections, this term has to be
properly subtracted in order to avoid double counting.  Given the
relevance of resummation for the large parameter $\alpha_s(\mu_R)
\log(\mu_f/m_b)$, it is often concluded that the $2\to 2$ process
gives the bulk of the cross-sections for the single production of
$H^\pm$. As was already noticed in the previous sections, however, the
issue of the dominance of one cross section over the other depends on
$\sqrt{s}$ and $m_{H^\pm}$.

\begin{figure}[p]
\begin{center}
\epsfxsize= 11.7cm
\leavevmode
\epsfbox[25 165 585 565]{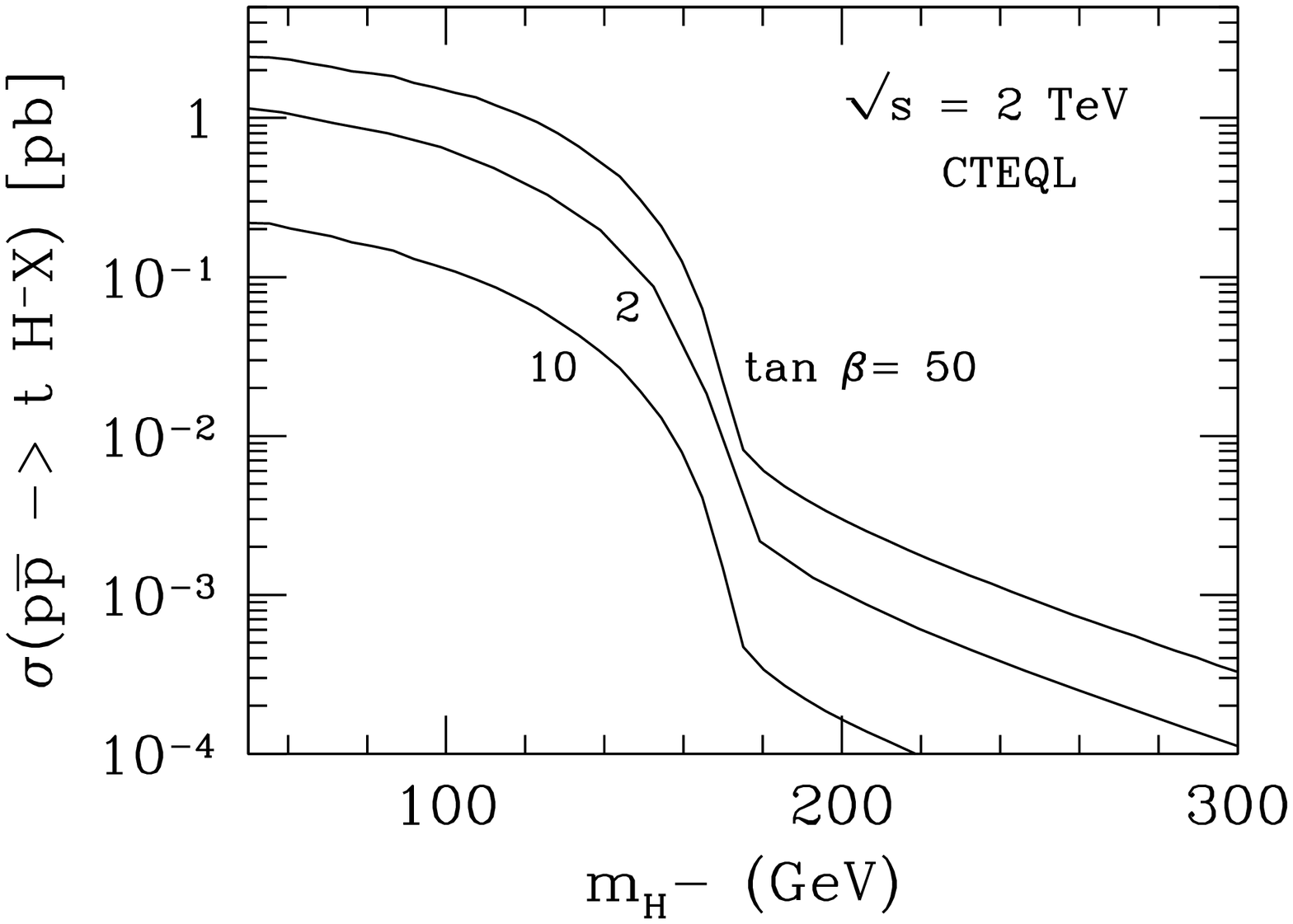}
\end{center}
\caption[f1]{The leading-order production cross section
 $\sigma(p\bar{p}\to t H^- X)$, for $\sqrt{s} = 2\,$TeV, as a function
 of the charged Higgs mass, is shown for three different values of
 $\tan \!\beta = 2,\,10,\,50$.  The cross section is obtained by
 adding the contribution of the $2 \to 2$ process, $g b \to t H^-$, to
 those of the $2 \to 3$ ones, $ g g \to t \bar{b} H^-$ and 
 $q \bar{q} \to t \bar{b} H^-$, and subtracting overlapping terms.
 Renormalization and factorization scales are fixed as
 $\mu_R\! =\! \mu_f\! =\! m_t+m_{H^-}$.}
\label{TeVsum}
\vspace*{0.5truecm}
\begin{center}
\epsfxsize= 11.7cm
\leavevmode
\epsfbox[25 165 585 585]{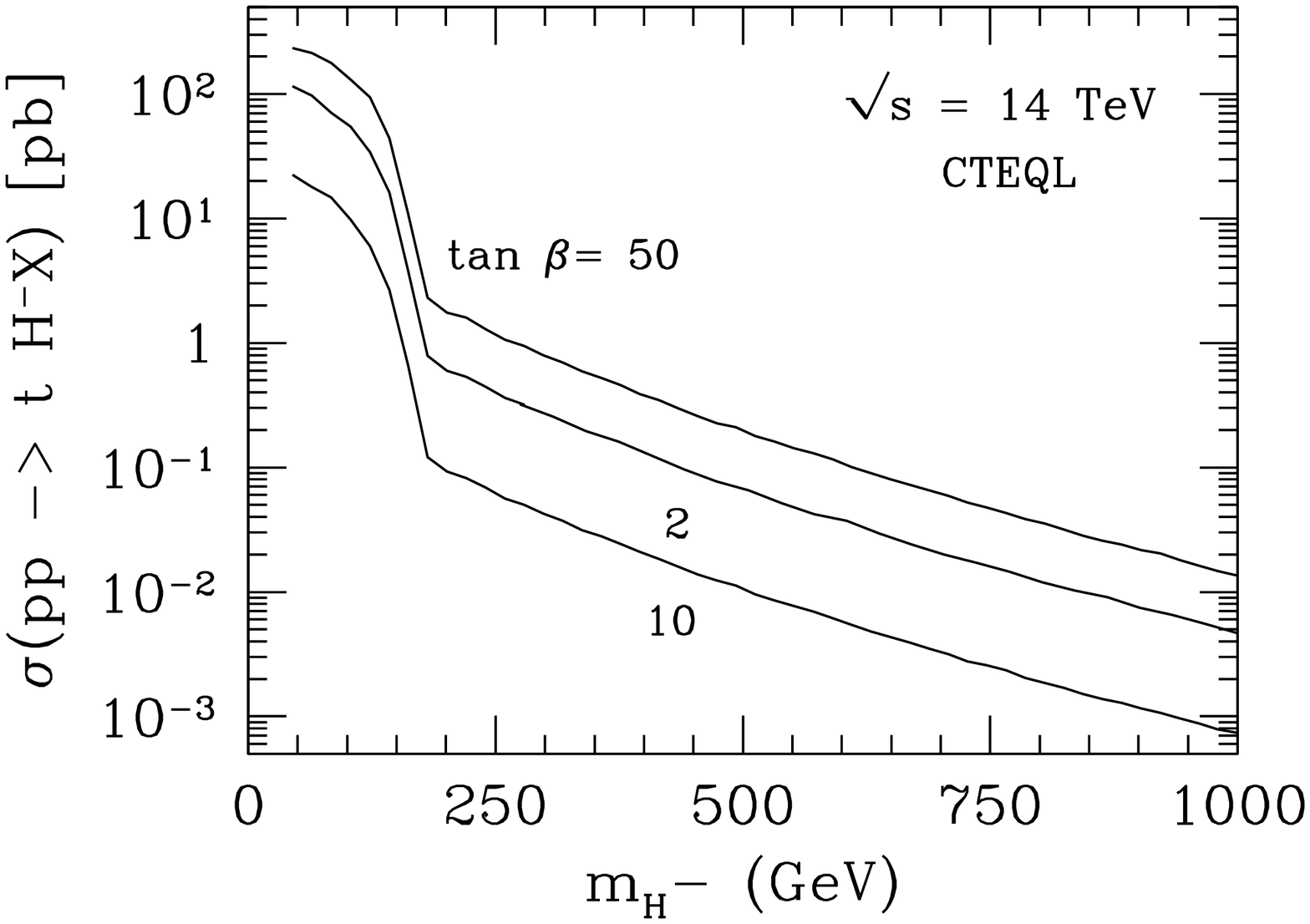}
\end{center}
\caption[f1]{Same as in Fig.~\ref{TeVsum} for center-of-mass energy
 $\sqrt{s}=14\,$TeV.}
\label{LHCsum}
\end{figure}

The summation and subtraction procedure has been carefully
systematized in the case of neutral Higgs bosons
production~\cite{NDOUBLECOUNTING,BBH}, for which, in a similar way,
also the $2\to 1$ process, $\bar{b} b \to H^0$, and the $2 \to 2$ one,
$g b\to b H^0$, partially overlap.  It was applied to the case of
charged-Higgs production in Ref.~\cite{GHPTW}, where it was debated
whether a $2\to1$ elementary process $t \bar{b} \to H^+$ also
contributes to the inclusive cross section, once the theoretically
calculated $t$-distribution function is folded in the proton beam.
For the ranges of $m_{H^\pm}$ that may be probed at the Tevatron and
at the LHC, the term $\log(m_H/m_t)$ is sufficiently small, and the
$2\to 1$ process can be safely omitted~\cite{GHPTW}.  We therefore
disregard the process $t\bar{b} \to H^+$ in our analysis and in
addition assume sufficient $b$-tagging efficiency so that final states
with three or more $b$'s can be distinguished~\cite{JGUNION,b-tag}.

{}Following  Refs.~\cite{NDOUBLECOUNTING,BBH}, 
we introduce the distribution function 
$\tilde b(x,\mu_f)$ given by the perturbative solution to the 
Dokshitzer--Gribov--Lipatov--Altarelli--Parisi equation:
\begin{equation}
\tilde b(x,\mu_f) = \frac{\alpha_s(\mu_R)}{\pi} 
\ln \left(\frac{\mu_f}{m_b}\right) 
\int^1_x \frac{dy}{y} P_{gb}\left(\frac{x}{y}\right) g(y,\mu_f) 
\end{equation}
where the splitting function is $P_{gb}(x) = (x^2+(1-x)^2)/2$ and
$g(x,\mu_f)$ is the usual gluon-distribution function at the
factorization scale $\mu_f$.  The hard process $g b\to t H^-$,
convoluted with the distribution function $\tilde b$ above, gives a
contribution that has to be subtracted from the sum of the
$gg$-initiated and (standard) $g b$-initiated process, convoluted with
the phenomenological $b$-distribution function.

The appropriately summed inclusive cross section is shown in
Fig.~\ref{TeVsum} for the Tevatron and in Fig.~\ref{LHCsum} for the
LHC.  As in the case of only the $2\to 3$ channels, a potential reach
of $\sim 250\,$GeV and of ${\cal{O}}(1)\,$TeV is found, respectively,
for the Tevatron (with $10$--$30\,$fb$^{-1}$ luminosity) and the LHC
(with a luminosity of $100\,$fb$^{-1}$ per year).  Background and
other detection and identification considerations will somewhat
diminish this reach~\cite{JGUNION}, which is therefore given only as a
rough guideline.

\section{Slepton-Strahlung}
\label{sec:s3}

In supersymmetric models, lepton $L$ and baryon $B$ numbers are not
accidental symmetries, but have to be imposed by hand, e.g. by
imposing a discrete $R$-symmetry ($R$-parity)~\cite{Rparity}
$R_{P}=(-1)^{3B+L+J}$, where $J$ is the particle spin.  It is
possible, however, that only $L$ or $B$ correspond to a conserved
number, which is sufficient to ensure the proton stability.  In
particular, in supersymmetry the lepton and Higgs doublets are not
distinguished by their spin as in the SM.  It is therefore natural to
expect that some mixing exists between slepton and Higgs bosons as
well as between leptons and higgsinos that carry the same quantum
numbers, and, hence, that the lepton number $L$ is generically not
conserved.  The realization of $L$-violation is basis-dependent and it
is usually convenient to define the two-Higgs doublets as those whose
neutral components are aligned along the two large vev's, whereas the
lepton doublets are along the orthogonal directions in field space.
In this basis it is straightforward to show that neutrino masses arise
from small tree-level mixing with the neutralinos and at one- and
two-loop levels from $\Delta L = 1$, Yukawa-type
interactions~\cite{NEUTRINO1,WIS,NEUTRINO2,NEUTRINO3}.  This offers an
exciting avenue for generating neutrino mass and mixing at the weak
scale. It further suggests collider tests of models of neutrino masses
since both radiative neutrino masses and slepton production are
controlled by the same Yukawa couplings.

In the following, only models in which the lepton number is violated by
$\Delta L = 1$ renormalizable operators are considered.  The
low-energy Lagrangian is derived from the superpotential operator
\begin{equation}
  W =\lambda^\prime_{lmn}L_{l}Q_{m}D_{n} \,,
\label{superpot}
\end{equation}
where $L,\,Q,\,D$ are the lepton and quark-doublet and down-singlet
superfields, respectively.  The possibility of renormalizable purely
leptonic operators, also involving a $\Delta L =1$ lepton-number
violation, does not affect our analysis, and such operators
are neglected hereafter.  In
component fields, the slepton--quark interactions relevant to our
purposes are:
\begin{equation}
{\cal L}_{\not L}  \ \supset \  
  \lambda^\prime_{lmn}  V_{mj} \, 
   {\overline{u}_L}_j \, {d_R}_n  \tilde{e}_{L\,l}^{\,\ast} 
\ - \ 
  \lambda^{\prime\,\ast}_{lmn} \,
   {\overline{d}_L}_m \, {d_R}_n  \tilde{\nu}_{L\,l}^{\,\ast}
                                              +{\rm h.c.}\,, 
\label{rpvlag}
\end{equation}
where $\tilde{f}$ denotes the sfermion superpartner of a fermion $f$
and all generation indices are arbitrary.  In deriving~(\ref{rpvlag})
it was assumed that the right-handed quark fields as well as the
left-handed down-quark fields are already in the mass eigenstate basis,
and that the CKM matrix $V$ coincides with the rotation matrix of the
left-handed up-quark sector.  Had this assumption not been made, both
fermions in each of the two terms of eq.~(\ref{rpvlag}) would be
multiplied by the appropriate rotation matrix.  However, in the
absence of any initial assumption on the texture of the
$\lambda^{\prime}$ matrix, this would merely correspond to a
redefinition of its elements.

The $\Delta L = 1$ couplings given above cannot be arbitrarily large
as they lead to tree-level corrections to various observables and
correct neutral- and charged-current universality~\cite{CONSTRAINTS}.
With the exception of $\lambda^\prime_{111} \lesssim 10^{-4}$ and
$\lambda^\prime_{1m1} \lesssim 10^{-2}$ from
$(\beta\beta)_{0\nu}$-decay and atomic parity violation, respectively,
one has $\lambda^\prime_{11n} \lesssim 0.01$, 
$\lambda^\prime_{l2n} \lesssim 0.2$, and 
$\lambda^\prime_{l3n} \lesssim 0.4$--$0.5$, for sfermions with masses of 
$100\,$GeV. These constraints and their derivation are summarized, for
instance, in Ref.~\cite{RPV}.  As an example, the weak constraints on
$\lambda^\prime_{l3n}$ are derived from either $t$-quark
decays~\cite{KAON,SNEUTRINO} (this constraint, however, vanishes as
the slepton mass approaches the $t$-quark mass), from the one standard
deviation in the $Z$ width~\cite{Z}, or from $b$-quark semileptonic
decays~\cite{SNEUTRINO,B}, again at one standard deviation.  The above
constraints scale as a power of $m_{\tilde{f}}$ and are therefore
significantly weaker for heavier sfermions.  For example, for squarks
near the 500 GeV mark, $\lambda^\prime \sim 1$ in the third family is
generally not excluded.  On the other hand, constraints on pairs of
non-identical couplings (e.g. from meson mixing) often imply, further,
that certain combinations of couplings cannot saturate their individual
upper bounds simultaneously.  In general, the hierarchy of couplings
that emerges from experiment is similar to the generational hierarchy
in the usual Yukawa couplings (with $\lambda^{\prime}_{333}$ the most
weakly constrained), an observation that we will adopt as a guideline.
(A similar structure is also suggested by various theoretical models.
See, for example, Refs.~\cite{NEUTRINO1,NEUTRINO2}.)

\subsection{Production cross section}

The terms~(\ref{rpvlag}) in the Lagrangian lead to new and exciting
possibilities for slepton production at hadron colliders: $(i)$ exotic
$t$-quark decays $t \to \tilde{\tau}b$, if kinematically
allowed~\cite{KAON,SNEUTRINO,TOP}, $(ii)$ $s$-channel resonant production of
sleptons~\cite{RESONANCE} (see, for instance, Ref.~\cite{SNEUTRINO} for a
discussion of resonant production at LEP), and, as proposed above,
$(iii)$ associate production $qq^{\prime}\tilde{l}$.  (In addition,
$(iv)$ gluonic couplings are induced for the sneutrino, which could now
be singly produced $gg\to\tilde{\nu}$~\cite{FUSION2}.)  Here,
we focus on the charged-slepton-strahlung production, in
particular,  stau $\tilde{\tau}$ production in association with $t$-
and $b$-quarks.  Associate production of the neutral sleptons will be
discussed elsewhere~\cite{FUSION2}.  Both $2 \to 2$ and $2 \to 3$
channels $gb \to t \tilde{\tau}^-$ and 
$gg,\, qq \to t \tilde{\tau}^{-} \bar{b}$, will be considered (at the
leading order) and properly combined.  As in the case of the charged
Higgs production, the $2 \to 3$ processes encompass the production
mechanism $(i)$ in the relevant kinematic region.

Concretely, the only $R$-parity-violating coupling that is
considered is $\lambda^{\prime}_{333}$.  This choice is motivated by
the fact that this is the least constrained coupling, and, as
explained above, it is expected to be the most significant among the
$\Delta L = 1$ couplings in some frameworks.  In this case, the
cross section scales as $\lambda^{\prime\,2}_{333}$ in the kinematic
region $m_{t} < m_{\tilde{\tau}} + m_{b}$. In the complementary region
$m_{t} \gtrsim m_{\tilde{\tau}} + m_{b}$, sleptons contribute to the
width of the $t$-quark (in proportion to $\lambda^{\prime\,2}_{333}$),
violating this simple scaling law. This implies that, although our
results may be taken as indicative for cases involving other
$\lambda^{\prime}$ couplings, they cannot be used directly in cases
not involving the $t$-quark.  In addition, the subtraction procedure
to be followed when combining the different $ 2 \to 2 $ and $2 \to 3$
channels into the inclusive cross section, although conceptually
similar, differs technically for cases with one or two light quarks
associated to the $\tilde{\tau}$-production.  For this reason also,
the results presented here cannot be simply adapted to production
cross sections involving other $\lambda^{\prime}$ couplings or to the
case of neutral-slepton production.  (The latter is also complicated
by the presence of the gluon-fusion channel.)  On the other hand, the
slepton generation label does not enter our calculation but affects
only the signal analysis (on which we comment below). Hence, our
results can be generalized in a straightforward fashion to the
production of other charged sleptons in association with $t$- and
$b$-quarks.

It should be noted that the produced slepton is always left-handed, as
a result of the structure of the operator~(\ref{rpvlag}).  Of course,
the physical eigenstates are, in general, admixtures of left- and
right-handed current eigenstates.  For simplicity, it is assumed in
the following that the left--right mixing term in the $\tilde{\tau}$
mass squared matrix is small. As a consequence, left- and right-handed
current eigenstates coincide already with the two mass eigenstates,
with masses $m_{\tilde{\tau}_L}$ and $m_{\tilde{\tau}_R}$. The first of
these two masses, recurrent in this analysis, will be simply denoted as
$m_{\tilde{\tau}}$. (A generalization is straightforward and involves
the introduction of a mixing angle.)

The inclusive production cross sections 
$p\bar{p} \to t \tilde{\tau}_{L} X$, $p {p} \to t \tilde{\tau}_{L} X$ 
are obtained by combining the production cross sections arising from
the $2 \to 2$ elementary process $ g b \to t \tilde{\tau}_L$ to those
induced by $2\to 3$ partonic processes, which give rise to 
$p\bar{p}\to t \bar{b} \tilde{\tau}_{L} X$, 
$p {p} \to t \bar{b}\tilde{\tau}_{L} X$. The $2 \to 3$ processes 
might be independently measured only if relatively complicated final
states could be detected.  The corresponding inclusive cross sections,
formally given by eq.~(\ref{Xsec}), is obtained by convoluting the
hard-scattering cross section of quark- and gluon-initiated processes
with the quark and gluon distribution functions in $p$ and
$\bar{p}$. The Feynman diagrams for the partonic processes are those
of Figs.~\ref{qdiags} and~\ref{gdiags}, with $H^-$ replaced by
$\tilde{\tau}_L$.  Since the vector and axial coupling $v$ and $a$ for
the vertex $t$--$b$--$\tilde{\tau}_L$ are now simply $v=a=1/2$, the
square amplitudes 
$\vert {\cal M} \vert^2_{q \bar{q}}$ and 
$\vert {\cal M} \vert^2_{gg}$ 
can be decomposed as:
\begin{eqnarray}
\vert {\cal M} \vert^2_{q \bar{q}}  & = & 
\frac{1}{4} \lambda^{\prime\,2}_{333} \left( 4\pi \alpha_S\right)^2 
 \vert V_{tb}\vert^2 
\left( V^{q \bar{q}}  +  A^{q \bar{q}}\right)
  \nonumber \\
\vert {\cal M} \vert^2_{gg}        & = &
\frac{1}{4} \lambda^{\prime\,2}_{333} \left( 4\pi \alpha_S\right)^2 
 \vert V_{tb}\vert^2 
\left( V^{gg} +  A^{gg} \right)\,.
\label{rpamplitdecomp}
\end{eqnarray}
The reduced square amplitudes $V^{q \bar{q}}$, $ A^{q \bar{q}}$ and
$V^{gg}$, $ A^{gg}$ coincide with those obtained for the $2 \to 3$
charged-Higgs-production processes, once the replacement 
$m_{H^\pm} \to m_{\tilde{\tau}_L}$ is made.  In the kinematical region
of a resonant $t$-quark, only the two decay modes $t\to W^+ b $ and 
${t} \to \tilde{\tau}^{+}{b}$ are considered. The latter has the width
\begin{equation}
 {\Gamma({t}\to \tilde{\tau}^{+}{b})  = } 
 \frac{\lambda^{\prime\,2}_{333}}{32 \pi} 
 \,m_t \vert V_{tb}^\ast \vert^2  
 \left( 1\!+\! \frac{m_b^2}{m_t^2} - 
               \frac{m_{\tilde{\tau}}^2}{m_t^2}
 \right) 
 \lambda^{1/2}\left(1,\frac{m_{\tilde{\tau}}^2}{m_t^2}, 
                               \frac{m_b^2}{m_t^2}\right)\,,
\label{stauwidth}
\end{equation}
{}For simplicity, the charged Higgs $H^\pm$ is assumed
to be sufficiently heavy, as to kinematically forbid the decay mode
${t}\to H^+ {b} $.

\begin{figure}[ht]
\begin{center}
\epsfxsize= 11.7cm
\leavevmode
\epsfbox[25 165 585 565]{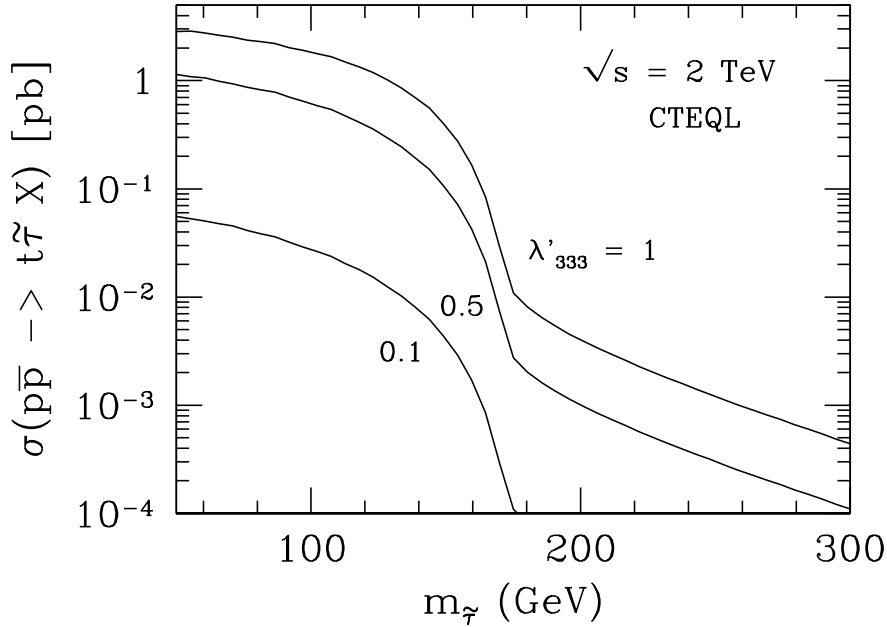}
\end{center}
\caption[f1]{The leading-order production cross section
 $\sigma(p\bar{p}\to t \tilde{\tau} X)$, for
 $\sqrt{s}=2\,$TeV, as a function of the $\tilde{\tau}$ mass, is
 shown  for different values of 
 $\lambda^\prime_{333}$. Renormalization and factorization scales
 are fixed as $\mu_R\! =\! \mu_f\! =\! m_t+m_{\tilde{\tau}}$.}
\label{TeVrviol}
\end{figure}

\begin{figure}[ht]
\begin{center}
\epsfxsize= 11.7cm
\leavevmode
\epsfbox[25 165 585 585]{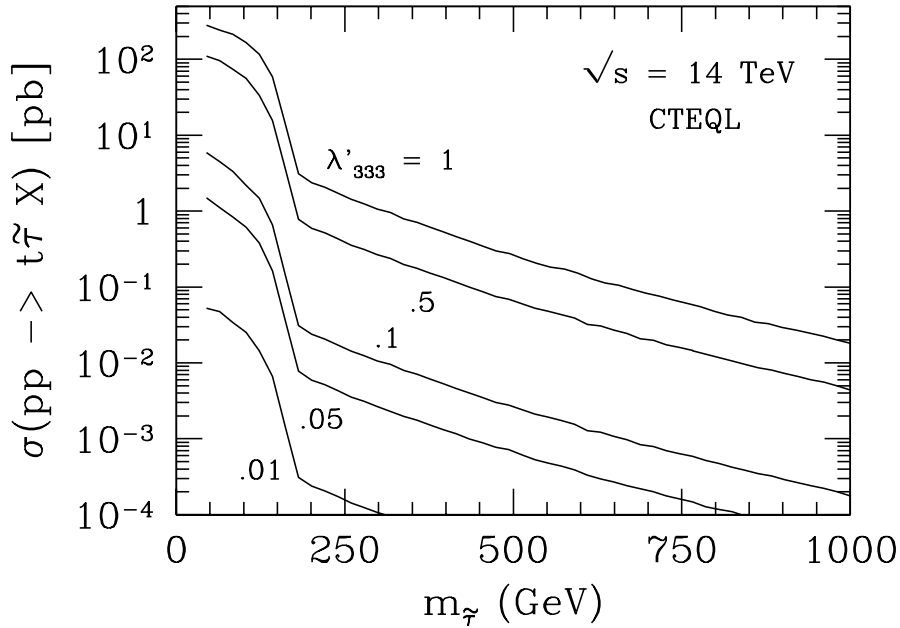}
\end{center}
\caption[f1]{The leading-order production cross section
 $\sigma(p p\to t \tilde{\tau} X)$, for
 $\sqrt{s}=14\,$TeV, as a function of the $\tilde{\tau}$ mass is
 shown for different values of 
 $\lambda^\prime_{333}$. Renormalization and factorization scales
are fixed as $\mu_R\! =\! \mu_f\! =\! m_t+m_{\tilde{\tau}}$.}
\label{LHCrviol}
\end{figure}

If only the $t$-quark can be detected in addition to the decay
products of $\tilde{\tau}_L$, the contribution from the $ 2 \to 2 $
process has to be included.  The Feynman diagrams for this process are
those of Fig.~\ref{gbdiags}, with the obvious substitution 
$H^- \to \tilde{\tau}_L$; the hard-scattering cross section is
obtained from~(\ref{siggb}) by replacing $4 {G_F} M_W^2 /{\sqrt{2}}$
with $\lambda^\prime_{333}/2$ and $m_{H^-}$ with $m_{\tilde{\tau}_L}$.
The subtraction procedure to avoid double counting follows exactly the
pattern already described for the production of the charged Higgs
boson.
The final inclusive production cross sections 
$\sigma(p\bar{p}, pp\to t\tilde{\tau} X)$ are shown in
Figs.~\ref{TeVrviol} and~\ref{LHCrviol} for the Tevatron and the LHC,
respectively, as a function of the $\tilde{\tau}_L$ mass
$m_{\tilde{\tau}_{L}} \gtrsim 45$ GeV.  Note that, since 
lepton number is violated, we conservatively apply the
model-independent lower limit $m_{\tilde{\tau}} \gtrsim 45$ GeV
extracted from the measurement of the $Z$ width.

The large cross section obtained in the case of $\sqrt{s} = 14\,$TeV
implies that at the LHC, with a luminosity of 100 fb$^{-1}$ per year,
light $\tilde{\tau}$'s may be produced in abundance even for couplings
as small as 0.01, whereas for large couplings, they may be produced up
to masses of ${\cal{O}}(1)\,$TeV.

The $\tilde\tau$-decay modes are highly model-dependent in the 
case of lepton-number violation. 
In particular, all superpartners typically decay in the collider 
and the typical large missing energy signature is replaced with 
multi-$b$ and lepton signatures, which may be used for 
identification. (See, for example, Ref.~\cite{RPTEV}.)
Depending on couplings and phase space, the main two-body 
decays for $\tilde{\tau}$ are, for example:
$\tilde{\tau} \to \tilde{\chi}^{0}\tau,\,
\tilde{\chi}^{\pm}\nu_{\tau},\,{t}b,\,{c}b, {t}s,\,l{\nu}$,  
where we included the effect of purely leptonic couplings.  (There
also exist two-body decays due to tree-level Higgs--slepton,
chargino--tau, and neutralino--neutrino mixing. However, these are
strongly suppressed by the small mixing angle 
$\sim (m_{\nu}/m_{Z})^{2}$.) In addition, various three-body decay
channels may be open, depending on the model parameters.
The charginos and neutralinos, if produced, also cascade in a
model-dependent way to leptons and jets.  Therefore, detection and
background studies cannot be done in a model- (coupling-) independent
fashion; particularly so, once our assumption that only the charged
Higgs or the stau (but not both) are produced in association with $t$
is generalized.  Nevertheless, many promising multilepton and
multi-$b$ signatures are available.  The study of detection aspects is
well motivated by the potential reach in small coupling and/or large
mass, but it is left for future works.

Leaving detection issues aside, in $L$-violating models, sleptons are
potentially more accessible (depending on the coupling) than in models
with lepton-number conservation, where their direct production relies
on the Drell--Yan process, which allows a discovery reach for sleptons
only up to $m_{\tilde{l}} \lesssim 350\,$GeV at the
LHC~\cite{SLEPTON}.  (Cascade decays may provide the bulk of slepton
production in the lepton-number-conserving models, but such processes
are highly model-dependent. They also complement slepton production in
the lepton-number-violating case studied here.)  The potential reach
at the Tevatron is limited to large couplings and/or light
$\tilde{\tau}$'s.  Nevertheless, the slepton-strahlung provides a
unique slepton-discovery mechanism at the Tevatron as well.

\subsection{Exploring $\Delta L = 1$ models of neutrino masses}

\begin{figure}[t]
\begin{center}
\epsfxsize= 6.0cm
\leavevmode
\epsfbox[200 565 420 685]{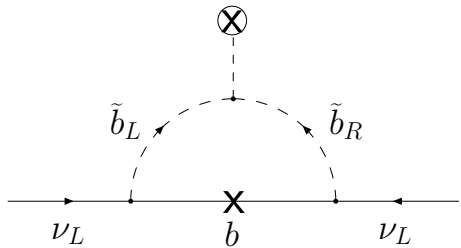}
\end{center}
\caption[f1]{A one-loop contribution to the Majorana neutrino mass 
 arising from the $\Delta L = 1$ operators of the
 superpotential~(\ref{superpot}).}
\label{neutrinomass}
\end{figure}

An obvious question is to know to what extent does the discovery reach
described above makes it possible to explore the corresponding models of
neutrino masses.  It is straightforward to show that the
supersymmetry-rotated operators~(\ref{rpvlag}) with a quark (sneutrino)
replaced by a squark (neutrino) lead, at one-loop order, to a Majorana
neutrino mass as illustrated in Fig.~\ref{neutrinomass}.
One obtains
\begin{equation}
\frac{m_{\nu}}{{\rm MeV}} 
\sim \lambda^{\prime\,2}
 \left(\frac{300\, {\rm GeV}}{m_{\tilde{b}}}\right)
 \left(\frac{m_{LR}^{2}}{m_{b}\,m_{\tilde{b}}}\right) \,,
\label{neumass}
\end{equation} 
where a $b$-quark and $\tilde{b}$-squark are assumed to circulate in
the loop, and $m_{LR}^{2}$ is the $\tilde{b}$ left--right mixing
squared mass.  (Note that the size $m_{LR}^{2}$ here can be
significant even if left--right stau mixing is suppressed.) The
neutrino mass may vanish in the limit of a continuous U(1)$_{R}$
symmetry, which corresponds in our case to 
$m_{LR}^{2} \ll m_{b}\,m_{\tilde{b}}$.  It is further assumed that no
other accidental cancellation among various contributions to the
neutrino mass takes place.  In this case, the
contribution~(\ref{neumass}), if it exists, constitutes probably the
dominant contribution to the neutrino mass (another contribution may
arise from tree-level neutrino--neutralino mixing).  Imposing
laboratory limits on the $\nu_{e}$ mass one can, for example, derive
severe constraints on $\lambda^\prime_{133}$~\cite{MNU}.

Given all other constraints and the cross sections of
Figs.~\ref{TeVrviol} and~\ref{LHCrviol}, $\lambda^\prime_{233}$ and
$\lambda^\prime_{333}$ are the couplings that are likely to be probed
at hadron colliders through slepton-strahlung production.  Assuming
that associate stau production can be observed for
$\lambda^{\prime}\gtrsim 0.1\, (0.01)$, neutrino masses heavier than
$3\,$KeV ($30\,$eV) can be explored, for 
$m_{\tilde{b}} \lesssim 1\,$TeV and $m_b\,m_{\tilde{b}}/ m_{LR}^2$ of
${\cal{O}}(1)$ in~(\ref{neumass}).  It is interesting to note that if
the coupling is large enough to lead to slepton production, it is
probable that the corresponding neutrino cannot be in the sub-eV
range, as is sometimes assumed, unless the $\tilde{b}$-squarks are in
the multi-TeV range, and with negligible left--right mixing.
Furthermore, a large slepton production cross section, may even imply
a heavy neutrino species, which decays on cosmological time scales.
Alternatively, if the $\tilde{b}$-squarks are not discovered at the
LHC (when considering the $R$-parity-violating cascades), the same
range of coupling would automatically imply much lighter neutrinos,
possibly in the sub-eV range, enhancing the coverage of the
neutrino-mass parameter space.

Thus, collider studies in this case carry indirect but crucial
implications to models of neutrino masses and can help reveal the
neutrino spectrum.  This beneficial relation is due to supersymmetry,
which relates the neutrino radiative mass operators and the
slepton--quark Yukawa operators.  Negative search results can
alternatively provide strong constraints on the
$\lambda_{i33}^{\prime}$ couplings, especially if slepton masses are
independently measured. Such potentially strong constraints are
currently not available by any other method.  Further handles on the
couplings and on the neutrino spectrum are provided by the search for
the sneutrinos~\cite{FUSION2}, in which case also the gluon fusion 
$gg \to \tilde{\nu}$ channel is available. The gluon-fusion cross
section also depends quadratically on $\lambda^{\prime}$ Yukawa
couplings, and it is given by a straightforward generalization of the
Higgs gluon-fusion $gg \to H^{0}$ cross section~\cite{FUSION2}.

\section{Summary and Outlook}
\label{sec:s4}

In summary, we have shown that charged Higgs- and slepton-strahlung
provide important channels of single production. Our study implies a
significant production cross section, in particular at the LHC,
but also at the upgraded Tevatron energies.  For example, 
charged Higgs bosons and staus as heavy as $1\,$TeV can be produced 
at the LHC (Figs.~\ref{LHCmassdep} and~\ref{LHCrviol}). For lighter 
staus, $R$-parity-violating couplings as small as 0.01 may be probed
(Fig.~\ref{LHCrviol}). At the Tevatron (Figs.~\ref{TeVsum}
and~\ref{TeVrviol}), the kinematic reach may be significantly
extended with respect to that obtained from the $t$-quark decays 
${t} \to H^{+}{b}$ and ${t} \to \tilde{\tau}^{+}{b}$. 
Of course, more conclusive statements should await 
detailed background and detection studies.

Our calculations, at the leading order, include exact evaluation of
the three-body phase space, and special attention was given to the
correct summation of the various contributions to the inclusive cross
section.

The importance of $b$-tagging in separating the various channels was
pointed out. In addition, various weakly interacting particles may be
produced simultaneously in association with quarks (e.g. the
charged Higgs and one or two charged sleptons), hence raising the
issue of their separation and identification.  In particular, in the
case of the stau studied here, the slepton signature could be similar
to that of the charged Higgs. (Similarly, the neutral Higgs and
sneutrino could both be produced via gluon fusion and/or strahlung and
could also decay similarly, extending the problem to the neutral
sector as well.) One may have to rely on mass differences and on more
suppressed $L$-violating or supersymmetric decays in order to
distinguish the different bosons.

Detection of singly produced sleptons via the strahlung process (or
any other process) carries substantial benefits to the mapping of the
lepton-number-violating potential (and superpotential).  Hence, it
also carries important consequences to models in which neutrino masses
are obtained radiatively, since radiative neutrino masses are
proportional to the couplings of the slepton--quark Yukawa operators.
Leaving detection issues aside, heavy sleptons could be abundantly
produced in $L$-violating models and small $L$-violating couplings (and
hence, small neutrino masses) may be probed.

Though not discussed explicitly, a similar situation to the one studied
in this paper can arise in any other model in which a weakly
interacting scalar couples via a non-negligible Yukawa coupling to
quarks. The most obvious example, which corresponds to a
straightforward generalization of the slepton-strahlung case, is
given by (scalar) lepto-quark models.

\acknowledgements
It is pleasure to thank H. Frisch and G. Moultaka for discussions.
This work was supported by CNRS and by the US Department of Energy
under contract No.~DE-FG02-96ER40559. NP wishes to thank the theory
group at CERN for its hospitality.

\newpage
\appendix
\section{Three-Body Phase Space}
\label{dphi3}

In eq.~(\ref{Xsec}) $d$PS $(q_1+q_2;p_1, p_2, p_3)$ is an element
of the 3-body final-state phase space normalized as~\cite{STOPS2}
\begin{eqnarray}
\int 
d {\rm PS} 
          &  =  &             
\int 
  (2 \pi)^4 \, \delta^4 (q_1\!+\!q_2\!-\!p_1\!-\!p_2-\!p_3) \,
  \frac{d^3 p_1}{(2\pi)^3 2 E_1} \,
  \frac{d^3 p_2}{(2\pi)^3 2 E_2} \,
  \frac{d^3 p_3}{(2\pi)^3 2 E_3}\,;         \nonumber \\[1.01ex]
          & \to & 
  \frac{1}{(2 \pi)^4}\frac{1}{8} \, 
\int_{-1}^{+1}  d (\cos\theta)
\int_{0}^{2\pi} d \phi 
\int_{E_1,\rm min}^{E_1,\rm max} d E_1 
\int_{E_{12}, \rm min}^{E_{12}, \rm max} d E_{12} \,,
\label{phsp}
\end{eqnarray}
where 
$q_{i}$ ($p_{i}$) are the four-momenta of the initial partons
(final-state particles) and
$E_1$, $E_2$ can be chosen, e.g. as the energies of the final $t$
and $b$ respectively, with $E_{12} \equiv (E_1-E_2)/2$, and $E_3$ the
energy of the $H^\pm$. The remaining integral in eq.~(\ref{phsp}) is
over appropriately defined angles $\theta$ and $\phi$, describing the
motion with respect to the beam axis of the three-momenta {\bf p}$_1$,
{\bf p}$_2$ of the two produced quarks.  (Of the initial four angular
integrations, one is eliminated from energy conservation and another
(azimuthal) angle integration simply gives a factor of $2\pi$ included
in eq.~(\ref{phsp}).) The integration bounds for the energy 
$E_1$ are as follows:
\begin{equation}
  E_{1,min} = m_t \,;
\hspace*{1truecm}
  E_{1,max} = \frac{\hat{s} + m_t^2 - \left(m_b +m_{H^\pm} \right)^2}
{2 \sqrt{\hat s}} \,;
\end{equation}
those for the energy $E_2$: 
\begin{equation}
  E_{12, min/max} = \frac{-b \mp \sqrt{b^2 - 4 a c }}{2 a } \,,
\end{equation}
with $a$, $b$, and $c$ given by:
\begin{eqnarray}
  a & = &  2 E_1 \sqrt{\hat{s}} - \hat{s} - m_t^2   
 \nonumber \\[1.1ex]
  b & = &  - \left(\sqrt{\hat{s}} - E_1 \right)
            \left(m_{H^\pm}^2 - m_b^2 \right)
 \nonumber \\
  c & = & \frac{1}{4} \left\{ \left(E_1^2 - m_t^2\right)
            \left[ \left( \sqrt{\hat{s}} -E_1\right)^2 - 4 m_b^2 \right]
          - \left(E_1^2 + m_{H^\pm}^2 -m_t^2 -m_b^2 \right)^2 \right\}\,.
\label{abc}
\end{eqnarray}

Finally, the resulting six-dimensional integral eq.~(\ref{Xsec})
over the remaining phase space and over the parton luminosities is
performed numerically with the standard Vegas Monte-Carlo integration
routine~\cite{VEGAS}.

\end{document}